\documentclass[a4paper]{article}
\usepackage{graphicx}
\usepackage{mathrsfs} 
\usepackage{amsmath,amssymb}

\def\bvec#1{\mbox{\boldmath $#1$}}

\begin{document}
\setlength{\baselineskip}{18pt}

\title{\bf Dissipative Dynamics of Entangled Finite-Spin Systems with Non-Competitive External Fields}


\author{Koichi Nakagawa \\[3mm] \small Hoshi University, Tokyo 142-8501, Japan}


\date{\empty}


\maketitle

\begin{abstract}
We apply a new method based upon thermofield dynamics (TFD) to study entanglement of finite-spin systems with non-competitive external fields for both equilibrium and non-equilibrium cases. For the equilibrium finite-spin systems, the temperature dependence of the extended density matrices is derived using this method, and the effect of non-competitive external field is demonstrated. For the non-equilibrium finite-spin systems, the time dependence of the extended density matrices and the extended entanglement entropies is derived in accordance with von Noumann equation, and the dissipative dynamics of the finite-spin systems is argued. Consequently, the applicability of the TFD-based method to describe entanglement is confirmed in both equilibrium and non-equilibrium cases with the external fieds.
\end{abstract}
\section{Introduction}
\label{sec:1}


Generally, entanglement states show a quantum-mechanically complicated behavior. A new method to analyze quantum entanglement using thermofield dynamics (TFD) \cite{Fano,Prigogine,Takahashi} has been proposed in Ref.~\cite{Hashizume}, and the capacity has been confirmed in Ref.~\cite{Nakagawa}. In this new treatment of quantum entanglement with TFD, an extended density matrix has been defined on the double Hilbert space (ordinary and tilde Hilbert spaces), and examined for some simple cases~\cite{Hashizume}. The new TFD-based method allows the entanglement states to be easily understood, because the intrinsic elements caused by quantum entanglement can be extracted from the extended density matrix in this formulation. Consequently, it has been found that the intrinsic quantum entanglement can be distinguished from the thermal fluctuations included in the definition of the ordinary entanglement at finite temperatures. Based on the analysis presented in Ref.~\cite{Hashizume}, it has been argued that the general TFD formulation of the extended density matrix is applicable not only to equilibrium states but also to non-equilibrium states. Furthermore, the extended entanglement entropy has been introduced using the extended density matrix, and compared with the traditional entanglement entropy for the case of non-equilibrium spin system without external field~\cite{Hashizume,Nakagawa}. However, analysis of the entanglement entropies of non-equilibrium systems with external fields has not been conducted in Refs.~\cite{Hashizume,Nakagawa} and, therefore, the examination of the entanglement entropies of non-equilibrium systems with external fields using TFD is of current interest. In the present communication, therefore, the extended density matrices and the entanglement entropies of finite-spin systems with and without the non-competitive external fields are exhaustively investigated in both the equilibrium and non-equilibrium cases, based upon the general TFD algorithm.

The rest of the paper is organized as follows. In the next section, we introduce the extended density matrix in the double Hilbert space of the TFD, and review the equilibrium systems with external fields. In Sec. 3, we obtain the the extended density matrices and the the extended entanglement entropies of non-equilibrium spin systems with and without the non-competitive external fields, and discuss the numerical results of the time dependence. Finally, conclusions
are described in Sec. 4.

\section{The extended density matrix metod and the example of equilibrium finite-spin systems }
\label{sec:2}

First of all, we give a short introduction of the extended density matrix, $\hat{\rho }$, in the double Hilbert space of the TFD. $\hat{\rho }$ has been defined in Ref. \cite{Hashizume} as follows: 
\begin{align} 
\hat{\rho }:=|\Psi \rangle\langle\Psi |,\ |\Psi \rangle:=\rho ^{1/2}\sum _{s}|s,\tilde{s}\rangle=\rho ^{1/2}\sum _{s}|s\rangle|\tilde{s}\rangle, \label{eq:12} 
\end{align} 
using the ordinary density matrix, $\rho $, where $\left\{ |s\rangle \right\}$ is the orthogonal complete set in the original Hilbert space and $\left\{ |\tilde{s}\rangle \right\}$ is the same set in the tilde Hilbert space of the TFD \cite{Suzuki1,Suzuki4}. If entanglement subsystems A and B are being examined, each of $|s\rangle$ and $|\tilde{s}\rangle$ states are represented as the direct products $|s_{\text{A}}, s_{\text{B}}\rangle=|s_{\text{A}}\rangle | s_{\text{B}}\rangle$ and $|\tilde{s}_{\text{A}},
\tilde{s}_{\text{B}}\rangle=|\tilde{s}_{\text{A}}\rangle |\tilde{s}_{\text{B}}\rangle$, respectively. Moreover, it has been proven in Ref. \cite{Hashizume} that if the square-root density matrix $\rho^{1/2}$ is written using the matrix elements $a_{\alpha_{\text{A}},\beta_{\text{B}},\alpha'_{\text{A}},\beta'_{\text{B}}}$ as 
\begin{equation} 
\rho^{1/2}:=
\sum_{\alpha_{\text{A}},\beta_{\text{B}},\alpha'_{\text{A}},\beta'_{\text{B}}}a_{\alpha_{\text{A}},\beta_{\text{B}},\alpha'_{\text{A}},\beta'_{\text{B}}}|\alpha_{\text{A}},\beta_{\text{B}}\rangle\langle \alpha'_{\text{A}},\beta'_{\text{B}}|,\label{eq:13} 
\end{equation} 
the renormalized extended density matrix $\hat{\rho }_{\text{A}}$ is then expressed as
\begin{align} 
\hat{\rho}_{\text{A}}&:={\text{Tr}}_{\text{B}}\hat{\rho}:=\sum_{\gamma_{\text{B}},\tilde{\gamma}'_{\text{B}}}\langle\gamma_{\text{B}},\tilde{\gamma}'_{\text{B}}|\hat{\rho}|\gamma_{\text{B}},\tilde{\gamma}'_{\text{B}}\rangle \notag \\
&=\sum_{\alpha_{\text{A}},\beta_{\text{A}},\alpha'_{\text{A}},\beta'_{\text{A}}} b_{\alpha_{\text{A}},\beta_{\text{A}},\alpha'_{\text{A}},\beta'_{\text{A}}} |\alpha_{\text{A}},\tilde{\beta}_{\text{A}}\rangle\langle\alpha'_{\text{A}},\tilde{\beta}'_{\text{A}}|\notag \\
&=\sum_{\alpha_{\text{A}},\beta_{\text{A}},\alpha'_{\text{A}},\beta'_{\text{A}}}
b_{\alpha_{\text{A}},\beta_{\text{A}},\alpha'_{\text{A}},\beta'_{\text{A}}}
\left(|\alpha_{\text{A}}\rangle\langle\alpha'_{\text{A}}|\right)\left(|\tilde{\beta}_{\text{A}}\rangle\langle\tilde{\beta}'_{\text{A}}|\right),
\label{eq:14b} 
\end{align} 
where 
\begin{equation}
b_{\alpha_{\text{A}},\beta_{\text{A}},\alpha'_{\text{A}},\beta'_{\text{A}}}=\sum_{\gamma_{\text{B}},\gamma'_{\text{B}}}
a_{\alpha_{\text{A}},\gamma_{\text{B}},\beta_{\text{A}},\gamma'_{\text{B}}}
a^{*}_{\alpha'_{\text{A}},\gamma_{\text{B}},\beta'_{\text{A}},\gamma'_{\text{B}}}.\label{eq:15b} 
\end{equation} 

Let us reexamine the $S=1/2$ spin system described by the Hamiltonian
\begin{align}
\mathcal{H}:=-J\bvec{S}_{\text{A}}\cdot \bvec{S}_{\text{B}}-\mu_BH\left(S_{\text{A}}^z+S_{\text{B}}^z\right) , 
\label{eq:1}
\end{align}
incorporating the spin operators, $\bvec{S}_{\text{A}}=(S_{\text{A}}^x, S_{\text{A}}^y, S_{\text{A}}^z)$ and $\bvec{S}_{\text{B}}=(S_{\text{B}}^x, S_{\text{B}}^y, S_{\text{B}}^z)$, of the subsystems, A and B, respectively, where $H$ is the magnitude of external field. The state, $|s\rangle$, of the total system is then denoted by the direct product, $|s\rangle=|s_{\text{A}},s_{\text{B}}\rangle=|s_{\text{A}}\rangle |s_{\text{B}}\rangle$. Using the base, $\left\{ |++\rangle, |+-\rangle, |-+\rangle, |--\rangle \right\} $, the matrix form of the Hamiltonian \eqref{eq:1} is then expressed as 
\begin{align}
\mathcal{H}&=\sum _{s_{\text{A}},s_{\text{B}},s'_{\text{A}},s'_{\text{B}}}
h_{s_{\text{A}},s_{\text{B}},s'_{\text{A}},s'_{\text{B}}}|s_{\text{A}},s_{\text{B}}\rangle \langle
s'_{\text{A}},s'_{\text{B}}|\nonumber \\ 
&=\left( -\frac{J}{4}-\mu_BH \right)|++\rangle\langle++|+\left( -\frac{J}{4}+\mu_BH \right)|--\rangle\langle--| \nonumber \\
&\hspace{3mm}+\frac{J}{4}\left( |+-\rangle\langle+-| +|-+\rangle\langle-+|\right)-\frac{J}{2}\left(
|+-\rangle\langle-+| +|-+\rangle\langle+-|\right) .\label{eq:1a}
\end{align}
As can be seen from Eq. \eqref{eq:1a}, the spin inversion symmetry of this system is broken by external field, $H$, in the Hamiltonian, $\mathcal{H}$. 

\begin{figure}
\begin{center}
\unitlength 1mm
\begin{picture}(150,90)
\put(0,53){
\scalebox{0.7}{
\put(0,0){\includegraphics{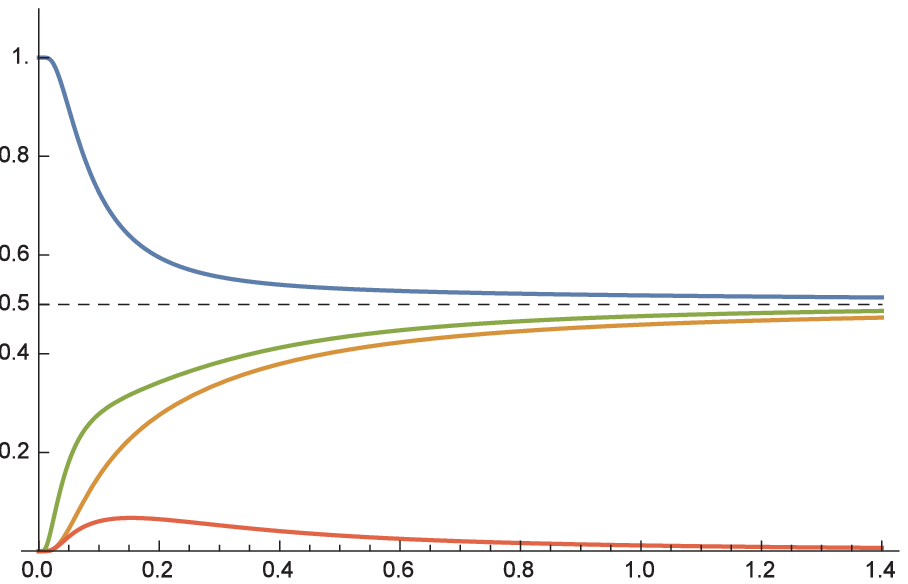}}
\put(94,.5){$(\beta J)^{-1}$}
\put(16,35){$b_{\rm d1}^{\rm eq}$}
\put(13,23){$b_{\rm d2}^{\rm eq}$}
\put(18,15){$b_{\rm cf}^{\rm eq}$}
\put(13,8){$b_{\text{qe}}^{\rm eq}$}

\put(47,-7){\scalebox{1.5}{(a)}}
}
}
\put(78,53){
\scalebox{0.7}{
\put(0,0){\includegraphics{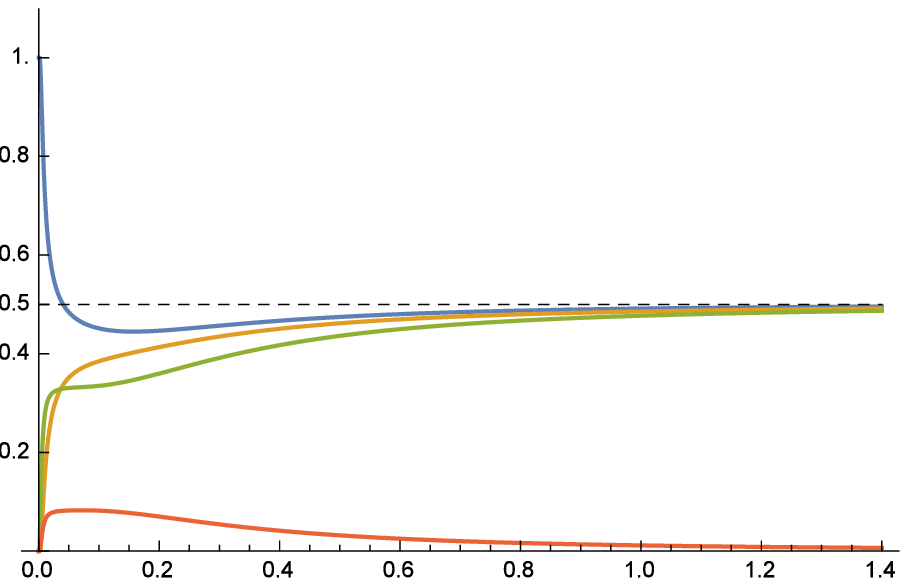}}
\put(94,.5){$(\beta J)^{-1}$}
\put(6,35){$b_{\rm d1}^{\rm eq}$}
\put(6,22){$b_{\rm d2}^{\rm eq}$}
\put(18,18){$b_{\rm cf}^{\rm eq}$}
\put(13,9){$b_{\text{qe}}^{\rm eq}$}

\put(47,-7){\scalebox{1.5}{(b)}}
}
}
\put(0,0){
\scalebox{0.7}{
\put(0,0){\includegraphics{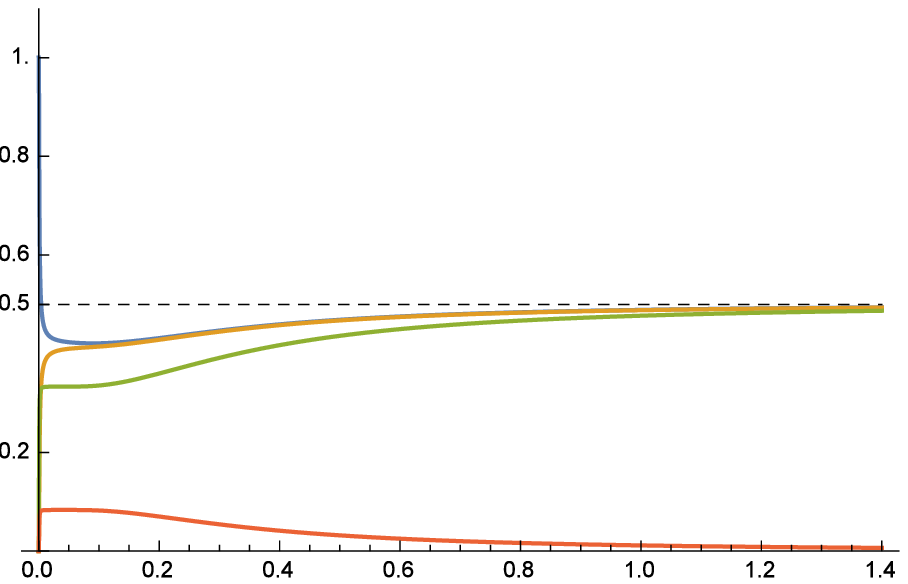}}
\put(94,.5){$(\beta J)^{-1}$}
\put(6,35){$b_{\rm d1}^{\rm eq}$}
\put(-2,22){$b_{\rm d2}^{\rm eq}$}
\put(18,18){$b_{\rm cf}^{\rm eq}$}
\put(13,9){$b_{\text{qe}}^{\rm eq}$}

\put(47,-7){\scalebox{1.5}{(c)}}
}
}
\put(78,0){
\scalebox{0.7}{
\put(1.5,0){\includegraphics{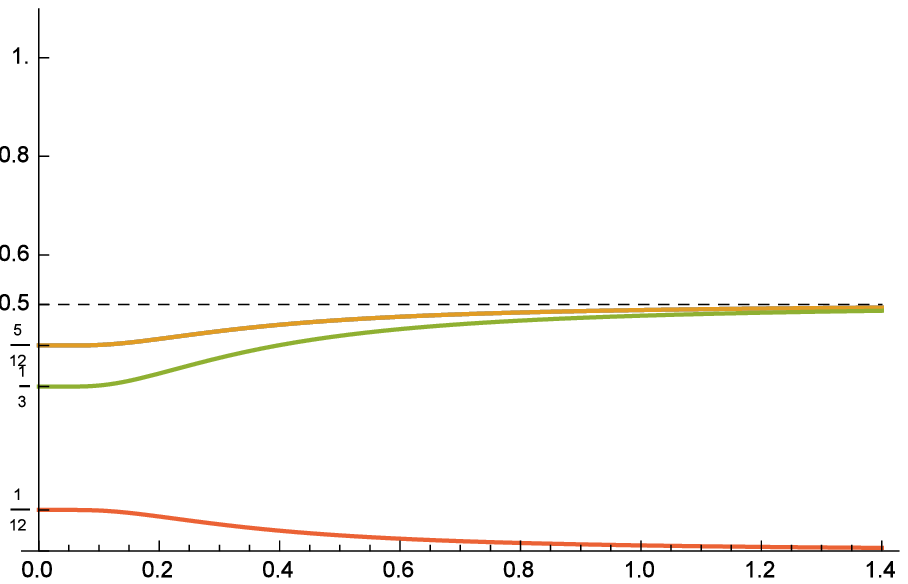}}
\put(95.5,.5){$(\beta J)^{-1}$}
\put(56,30){$b_{\rm d1}^{\rm eq}=b_{\rm d2}^{\rm eq}$}
\put(18,18){$b_{\rm cf}^{\rm eq}$}
\put(13,9){$b_{\text{qe}}^{\rm eq}$}

\put(47,-7){\scalebox{1.5}{(d)}}
}
}
\end{picture}
\end{center}
\caption{Temperature dependence of the elements $b_{\rm d1}^{\rm eq}, b_{\rm d2}^{\rm eq}, b_{\rm cf}^{\rm eq}$, and $b_{\rm qe}^{\rm eq}$ in Eqs. \eqref{eq:8}, $\sim $, \eqref{eq:11b}. Parts (a), (b), (c), and (d) show cases with $\mu_B H/J=0.1,~0.01, 0.001$ and $0$, respectively. The dashed lines in all the parts  represent the asymptotes of the $b_{\rm d1}^{\rm eq}, b_{\rm d2}^{\rm eq}$, and $b_{\rm cf}^{\rm eq}$ curves, and the asymptote of the $b_{\rm qe}^{\rm eq}$ curve is the holizontal axis. In parts (a), $\sim$, (c), at $\beta \to \infty $, $b_{\rm d2}^{\rm eq}, b_{\rm cf}^{\rm eq}$, and $b_{\rm qe}^{\rm eq}$ converge to the same value $0$, on the other hand, $b_{\rm d1}^{\rm eq}$ converges to 1. Part (d) shows that $b_{\rm d1}^{\rm eq}=b_{\rm d2}^{\rm eq}, b_{\rm cf}^{\rm eq}$, and $b_{\rm qe}^{\rm eq}$ converge to $\frac{5}{12},~\frac{1}{\,3\,}$ and $\frac{1}{12}$, respectively, at $\beta \to \infty $. }\label{fig:0}
\end{figure}

For the equilibrium states in terms of $\mathcal{H}$ expressed in Eq.~\eqref{eq:1a} , the ordinary density matrix, $\rho _{\rm eq}:=e^{-\beta \mathcal{H}}/Z(\beta)$, of this system can be obtained as
\begin{align}
\rho_{\text{eq}}&=\frac{e^{-K /4}}{Z(\beta)} \left( e^{K /2+h}|++\rangle\langle++| + e^{K /2-h}|--\rangle\langle--|\rule{0mm}{6mm}\right. \nonumber \\ 
&\hspace{16mm}+\cosh \frac{K}{2}\left(|+-\rangle\langle+-|+|-+\rangle\langle-+| \right) \nonumber \\ 
&\hspace{16mm}\left. +\sinh \frac{K}{2}\left(|-+\rangle\langle+-|+|+-\rangle\langle-+| \right) \right) ,\label{eq:1b}
\end{align}
where $\beta $ is the inverse temperature, the partition function, $Z(\beta):={\text{Tr}}e^{-\beta\mathcal{H}}$, is given by
\begin{equation}
Z(\beta)=2e^{-K/4}\left( e^{K/2}\cosh h+\cosh \frac{K}{2} \right),\label{eq:1c}
\end{equation}
$K:=\beta J$ and $h:=\beta \mu_B H$, respectively. For $\rho _{\rm eq}$ in Eq. \eqref{eq:1b}, the square-root density matrix, ${\rho _{\rm eq}}^{1/2}$, is then expressed as
\begin{align}
{\rho _{\rm eq}}^{1/2}&=\frac{e^h}{\sqrt{e^h \left(e^h+e^{-K}+1\right)+1}}~|++\rangle\langle++|\nonumber \\
&+\frac{1}{\sqrt{e^h \left(e^h+e^{-K}+1\right)+1}}~|--\rangle\langle--| \nonumber \\
&+\frac{e^{h/2}}{2}\left( \frac{1}{\sqrt{e^{2h}+e^h+1+e^{h-K}}} + \frac{1}{\sqrt{e^h+e^K\left( e^{2h}+e^h+1 \right)}}\right)\nonumber \\
&\hspace{10mm}\times \left( |+-\rangle\langle+-| +|-+\rangle\langle-+|\right) \nonumber \\ 
&+\frac{e^{h/2}}{2}\left( \frac{1}{\sqrt{e^{2h}+e^h+1+e^{h-K}}} - \frac{1}{\sqrt{e^h+e^K\left( e^{2h}+e^h+1 \right)}}\right)\nonumber \\ 
&\hspace{10mm}\times \left(|+-\rangle\langle-+| +|-+\rangle\langle+-|\right).\label{eq:5a}
\end{align}
It is of course that ${\rho _{\rm eq}}^{1/2}$ in Eq. \eqref{eq:5a} satisfies $\left( {\rho _{\rm eq}}^{1/2} \right) ^2=\rho _{\rm eq}$. According to formulae \eqref{eq:14b} and \eqref{eq:15b}, we then obtain the extended density matrix $\hat{\rho}_{\text{A}}^{\rm eq} :={\text{Tr}}_{\text{B}}\hat{\rho}^{\rm eq}:=\sum_{s_{\text{B}},\tilde{s}'_{\text{B}}}\langle s_{\text{B}},\tilde{s}'_{\text{B}}|\hat{\rho}^{\rm eq}|s_{\text{B}},\tilde{s}'_{\text{B}}\rangle$ of the spin A,
\begin{align}
\hat{\rho }_{\rm A}^{\rm eq}&=b_{\rm d1}^{\rm eq}|+\rangle\langle+||\tilde{+}\rangle\langle\tilde{+}|+b_{\rm d2}^{\rm eq}|-\rangle\langle-||\tilde{-}\rangle\langle\tilde{-}| \nonumber \\
&+b_{\rm cf}^{\rm eq}\left( |+\rangle\langle-||\tilde{+}\rangle\langle\tilde{-}| +|-\rangle\langle+||\tilde{-}\rangle\langle\tilde{+}| \right) \nonumber \\
&+b_{\rm qe}^{\rm eq}\left( |+\rangle\langle+||\tilde{-}\rangle\langle\tilde{-}| +|-\rangle\langle-||\tilde{+}\rangle\langle\tilde{+}| \right), \label{eq:6}
\end{align}
for the equilibrium system, where the matrix elements $b_{\rm d1}^{\rm eq}, b_{\rm d2}^{\rm eq}, b_{\rm cf}^{\rm eq}$ and $b_{\rm qe}^{\rm eq}$ are respectively obtained as analytic functions
\begin{align}
b_{\rm d1}^{\rm eq}&=\frac{e^{h} \left(4 e^{h+K}+\left(e^{K/2}+1\right)^2\right)}{4 \left(e^h+e^K\left( e^{2h}+e^h+1 \right)\right)}, \label{eq:8}\\
b_{\rm d2}^{\rm eq}&=\frac{\left(e^{K/2}+1\right)^2 e^h+4 e^K}{4 \left(e^h+e^K\left( e^{2h}+e^h+1 \right)\right)}, \label{eq:9}\\
b_{\rm cf}^{\rm eq}&=\frac{\cosh \left(\frac{K}{4}\right) \left(e^h+1\right) e^{\left(2 h+3 K\right)/4}}{e^h+e^K\left( e^{2h}+e^h+1 \right)}, \label{eq:10}
\end{align}
and
\begin{align}
b_{\rm qe}^{\rm eq}&=\frac{\sinh ^2\left(\frac{K}{4}\right)}{2 \cosh \left(\frac{K}{2}\right) \left(\cosh \left(h\right)+1\right)+2 \sinh \left(\frac{K}{2}\right) \cosh \left(h\right)} \label{eq:11b}
\end{align}
of $\beta$ and $H$, and correspond to the two diagonal elements (d1 and d2), the classical and thermal fluctuations (cf) and the quantum entanglements (qe) of $\hat{\rho}_{\text{A}}^{\rm eq}$, respectively. 
The numerical results of these matrix elements at $\mu _BH/J=0$ and $0.3$ have been obtained in Ref. \cite{Hashizume}, and the breaking of the spin inversion symmetry caused by the external field, $H>0$, has been argued, i.e. $b_{\rm d1}^{\rm eq} \neq b_{\rm d2}^{\rm eq}$ at $H>0$, and $b_{\rm d1}^{\rm eq} = b_{\rm d2}^{\rm eq}$ at $H=0$.
The parameter, $b_{\text{qe}}^{\rm eq}$, in Eq.~\eqref{eq:11b} expresses the quantum entanglement effect.
This quantum fluctuation is essential to quantum systems, and it has been used as an order parameter of 2D quantum systems \cite{Stephan,Tanaka}. As can be seen from Eq.~\eqref{eq:6}, only the intrinsic quantum entanglement is extracted clearly in the TFD formulation.
In particular, it can be understood that the entangled state of the system emerges through a single product, such as $|+\rangle\langle +||\tilde{-}\rangle\langle \tilde{-}|$, in $\hat{\rho}_{\text{A}}$. Let us now parenthetically touch upon the physical significance of $b_{\rm d1}^{\rm eq}, b_{\rm d2}^{\rm eq}, b_{\rm cf}^{\rm eq}$ and $b_{\rm qe}^{\rm eq}$. As can evidentlly be seen from Eqs. \eqref{eq:8}, $\sim $, \eqref{eq:11b}, the high temperature limits of $b_{\rm d1}^{\rm eq}, b_{\rm d2}^{\rm eq}, b_{\rm cf}^{\rm eq}$ and $b_{\rm qe}^{\rm eq}$ are
\begin{align}
\lim _{\beta \to 0}b_{\rm d1}^{\rm eq}=\lim _{\beta \to 0}b_{\rm d2}^{\rm eq}=\lim _{\beta \to 0}b_{\rm cf}^{\rm eq}=\frac{1}{\,2\,} ~~\mbox{and}~~\lim _{\beta \to 0}b_{\rm qe}^{\rm eq}=0, \label{eq:12a}
\end{align}
respectively, for any value of $H$. 
On the other hand, the low temperature limits of $b_{\rm d1}^{\rm eq}, b_{\rm d2}^{\rm eq}, b_{\rm cf}^{\rm eq}$ and $b_{\rm qe}^{\rm eq}$ are 
\begin{align}
\lim _{\beta \to \infty }b_{\rm d1}^{\rm eq}=1,~\lim _{\beta \to \infty }b_{\rm d2}^{\rm eq}=\lim _{\beta \to \infty }b_{\rm cf}^{\rm eq}=0 ~~\mbox{and}~~\lim _{\beta \to \infty }b_{\rm qe}^{\rm eq}=0, \label{eq:13a}
\end{align}
respectively, for $H>0$, and 
\begin{align}
\lim _{\beta \to \infty }b_{\rm d1}^{\rm eq}=\lim _{\beta \to \infty }b_{\rm d2}^{\rm eq}=\frac{5}{12},~\lim _{\beta \to \infty }b_{\rm cf}^{\rm eq}=\frac{1}{\,3\,} ~~\mbox{and}~~\lim _{\beta \to \infty }b_{\rm qe}^{\rm eq}=\frac{1}{12}, \label{eq:14}
\end{align}
respectively, for $H=0$. 
These results are summarized in Fig. \ref{fig:0}, and clarify that a transition from broken phase $(H>0)$ to unbroken phase $(H=0)$ of spin inversion symmetry happen in the equilibrium entanglement finite-spin system. 

\section{Non-equilibrium finite-spin systems with non-competitive external fields}
\label{sec:3}

Let us consider a non-equilibrium system with dissipation, which is described by the Hamiltonian of Eq.~\eqref{eq:1a}. We assume that the time dependence of the ordinary density matrix, $\rho (t)$, of this system is given by the dissipative von Neumann equation \cite{Suzukibook,Suzuki1}, which is described as
\begin{equation}
i\hbar\frac{\partial}{\partial t}\rho(t)=[\mathcal{H},\rho(t)]-\epsilon \left( \rho(t)-\rho_{\text{eq}}
\right),\label{eq:2}
\end{equation}
with $\epsilon $ being a dissipation parameter. The solution of Eq.~\eqref{eq:2} is then expressed as
\begin{equation}
\rho(t)=e^{-\epsilon t}U^{\dagger}(t)\rho_0 U(t)+(1-e^{-\epsilon t})\rho_{\text{eq}},\label{eq:3}
\end{equation}
for any initial condition, $\rho(0)=\rho_0$, where the unitary operator, $U(t):=e^{i\mathcal{H}t/\hbar}$, denotes
\begin{align}
U(t) &=e^{i\omega t/4} \left( \exp{\left(\frac{-i(\omega +2\mu_BH/\hbar)t}{2}\right)}|++\rangle\langle++| \right.\nonumber \\
&\hspace{13.5mm}+\exp{\left(\frac{-i(\omega -2\mu_BH/\hbar)t}{2}\right)}|--\rangle\langle--| \nonumber \\ 
&\hspace{13.5mm}+\cos \frac{\omega t }{2}\left( |+-\rangle\langle+-|+|-+\rangle\langle-+| \right) \nonumber \\ 
&\hspace{13.5mm}-\left.i\sin \frac{\omega t }{2}\left( |-+\rangle\langle+-|+|+-\rangle\langle-+| \right) \right), \label{eq:7}
\end{align}
 and $\omega:= J/\hbar$. The explicit expression of $\rho(t)$ in Eq.~\eqref{eq:3} is complicated for any initial condition, so hereafter, we confine ourselves to the initial condition, $\rho_0=|+-\rangle\langle+-|$. The insertion of Eqs.~\eqref{eq:1b} and \eqref{eq:7}, along with the initial condition, into Eq.~\eqref{eq:3}, brings forth
\begin{align}
\rho(t)=\frac{e^{-\epsilon t }}{2}&\left(\frac{2 \left(e^{\epsilon t }-1\right) e^{K+2 h}}{e^K\left( e^{2h}+e^h+1 \right)+e^h} |++\rangle\langle++|\right.\nonumber \\ 
&+\frac{2 \left(e^{\epsilon t }-1\right) e^{K}}{e^K\left( e^{2h}+e^h+1 \right)+e^h}|--\rangle\langle--| \nonumber \\ 
&+\left( \frac{\left(e^{\epsilon t }-1\right) \cosh \left(\frac{K}{2}\right)}{\cosh \left(\frac{K}{2}\right)+\cosh (h) e^{K/2}}+\cos \omega t +1 \right)|+-\rangle\langle+-|\nonumber \\ 
&+\left( \frac{\left(e^{\epsilon t }-1\right) \cosh \left(\frac{K}{2}\right)}{\cosh \left(\frac{K}{2}\right)+\cosh (h) e^{K/2}}-\cos \omega t +1 \right)|-+\rangle\langle-+|\nonumber \\ &+\left( \frac{e^h \left(e^{\epsilon t }-1\right) \left(e^{K}-1\right)}{e^K\left( e^{2h}+e^h+1 \right)+e^h}-i \sin \omega t  \right)|+-\rangle\langle-+|\nonumber \\ &\left.+\left( \frac{e^h \left(e^{\epsilon t }-1\right) \left(e^{K}-1\right)}{e^K\left( e^{2h}+e^h+1 \right)+e^h}+i \sin \omega t  \right)|-+\rangle\langle+-| \right) ,\label{eq:08}
\end{align}
where $h:=\mu_B \beta H/(\omega \hbar )$. 

The ordinary entanglement (von Neumann) entropy, $S:=-k_{\text{B}}{\text{Tr}}_{\text{A}}\left[ \rho_{\text{A}}\log \rho_{\text{A}} \right] $ and $\rho_{\text{A}}:={\text{Tr}}_{\text{B}}\rho(t)$, are then paraphrased into
\begin{align}
&S=-\frac{k_{\text{B}}}{2} e^{-t \epsilon } \left( \left(\frac{\left(\left(e^h+2\right) e^K+e^h\right) e^{t \epsilon }+\left(e^{2 h}-1\right) e^K}{\left(e^h+e^{2 h}+1\right) e^K+e^h}-\cos \omega t \right) \right. \nonumber \\
&\times \log \left(\frac{e^{-t \epsilon } }{2} \left(\frac{\left(\left(e^h+2\right) e^K+e^h\right) e^{t \epsilon }+\left(e^{2 h}-1\right) e^K}{\left(e^h+e^{2 h}+1\right) e^K+e^h}-\cos \omega t \right)\right) \nonumber \\
&+\left(\frac{\left(2 e^{h+K}+e^K+1\right) e^{h+t \epsilon }-\left(e^{2 h}-1\right) e^K}{\left(e^h+e^{2 h}+1\right) e^K+e^h}+\cos \omega t \right) \nonumber \\
&\times \log \left(\frac{e^{-t \epsilon } }{2}\left(\frac{\left(2 e^{h+K}+e^K+1\right) e^{h+t \epsilon }-\left(e^{2 h}-1\right) e^K}{\left(e^h+e^{2 h}+1\right) e^K+e^h}+\cos \omega t \right)\right). \label{eq:11}
\end{align}
It is also worth mentioning that $S$ in Eq.~\eqref{eq:11} is proportional to an entanglement, $E(C)$, which is a function of the ``concurrence''\cite{Wootters}, 
\begin{align}
C&=\frac{e^{-\epsilon t }}{\sqrt{2} \left(e^K\left( e^{2h}+e^h+1 \right)+e^h\right)}\nonumber \\
&\hspace{-5mm}\times \left( \left(-\left(2 e^{h+K}+e^K+1\right) e^{e t+h}-\left(\left(e^h+e^{2 h}+1\right) e^K+e^h\right) \cos (t \omega )+\left(e^{2 h}-1\right) e^K\right)\right.\nonumber \\
&\hspace{-5mm}\left.\times  \left(-e^{e t} \left(\left(e^h+2\right) e^K+e^h\right)+\left(\left(e^h+e^{2 h}+1\right) e^K+e^h\right) \cos (t \omega )-\left(e^{2 h}-1\right) e^K\right) \right)^{1/2}.
 \label{eq:11a}
\end{align}
The time dependence of $S$ and $C$ is displayed in Fig.~\ref{fig:02} (in units of $k_{\text{B}}=1$). In the dissipative system, at $t\to \infty $, $S$ and $C$ converge to the values, 
\begin{align}
-\frac{2 e^{K}+e^{K+h}+e^h}{2 \left(e^K\left( e^{2h}+e^h+1 \right)+e^h\right)}\log \left(\frac{2 e^{K}+e^{K+h}+e^h}{2 \left(e^K\left( e^{2h}+e^h+1 \right)+e^h\right)}\right)\nonumber \\
-\frac{e^h \left(e^{K}+2 e^{K+h}+1\right) }{2 \left(e^K\left( e^{2h}+e^h+1 \right)+e^h\right)}\log \left(\frac{e^h \left(e^{K}+2 e^{K+h}+1\right)}{2 \left(e^K\left( e^{2h}+e^h+1 \right)+e^h\right)}\right),
 \label{eq:12b}
\end{align}
and 
\begin{align}
\frac{\sqrt{2e^{K+2 h} (2 \sinh K+(2 \cosh (h)+3) \cosh K+2 \cosh (h) (\sinh K+1)+1)}}{e^K\left( e^{2h}+e^h+1 \right)+e^h},
 \label{eq:15}
\end{align}
respectively, so it is possible to argue that $S$ and $C$ include not only the contribution of the quantum entanglement, but also the contribution of the classical and thermal fluctuations in the equilibrium case. However, this fact is not  clarified in Eqs. \eqref{eq:11} and \eqref{eq:11a}.

\begin{figure}
\begin{center}
\unitlength 1mm
\begin{picture}(150,80)
\put(0,45){
\scalebox{0.7}{
\put(0,0){\includegraphics{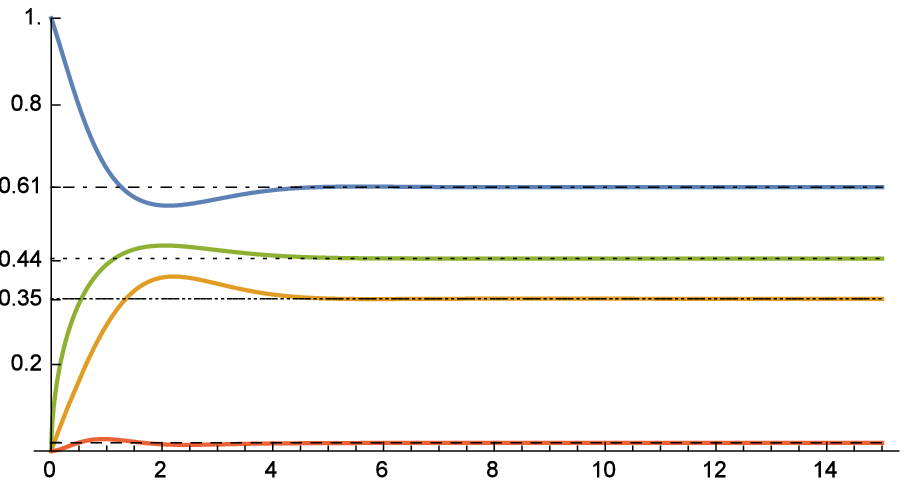}}
\put(94,.5){$\omega t$}
\put(7,44){$b_{\text{d1}}^{\rm neq}$}
\put(30,25){$b_{\text{cf}}^{\rm neq}$}
\put(9,10){$b_{\rm d2}^{\rm neq}$}
\put(18,6){$b_{\rm qe}^{\rm neq}$}
\put(47,-7){\scalebox{1.5}{(a)}}
}
}
\put(78,45){
\scalebox{0.7}{
\put(0,0){\includegraphics{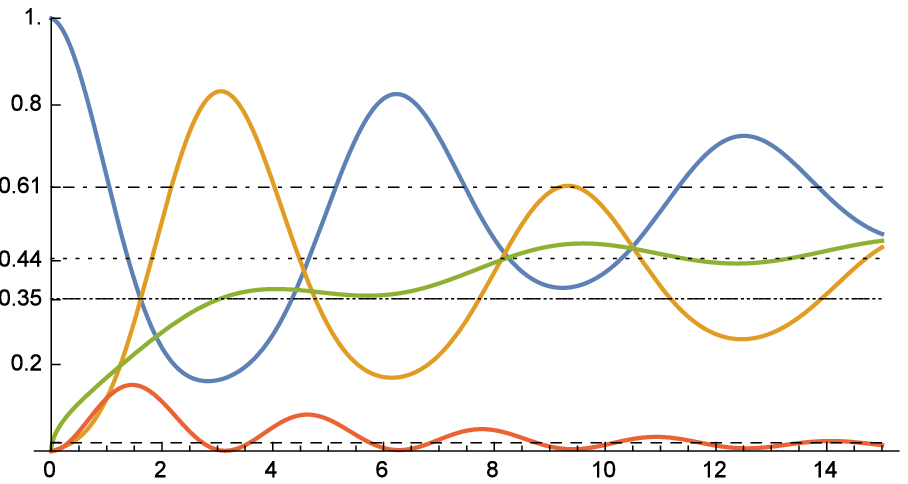}}
\put(94,.5){$\omega t$}
\put(7.5,44){$b_{\text{d1}}^{\rm neq}$}
\put(18,21){$b_{\text{cf}}^{\rm neq}$}
\put(25,38){$b_{\rm d2}^{\rm neq}$}
\put(19,6){$b_{\rm qe}^{\rm neq}$}
\put(47,-7){\scalebox{1.5}{(b)}}
}
}
\put(0,0){
\scalebox{0.7}{
\put(0,0){\includegraphics{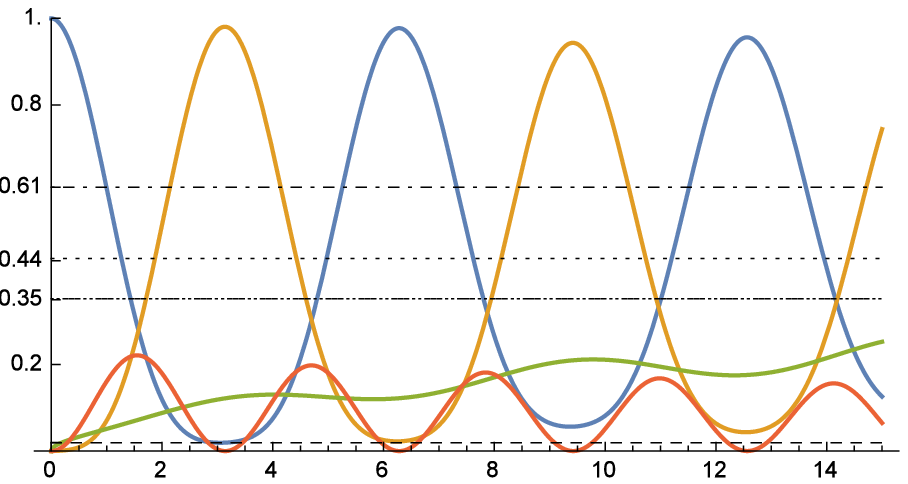}}
\put(94,.5){$\omega t$}
\put(8,44){$b_{\text{d1}}^{\rm neq}$}
\put(38,10.5){$b_{\text{cf}}^{\rm neq}$}
\put(25,44){$b_{\rm d2}^{\rm neq}$}
\put(17,11){$b_{\rm qe}^{\rm neq}$}
\put(47,-7){\scalebox{1.5}{(c)}}
}
}
\put(78,0){
\scalebox{0.7}{
\put(0,0){\includegraphics{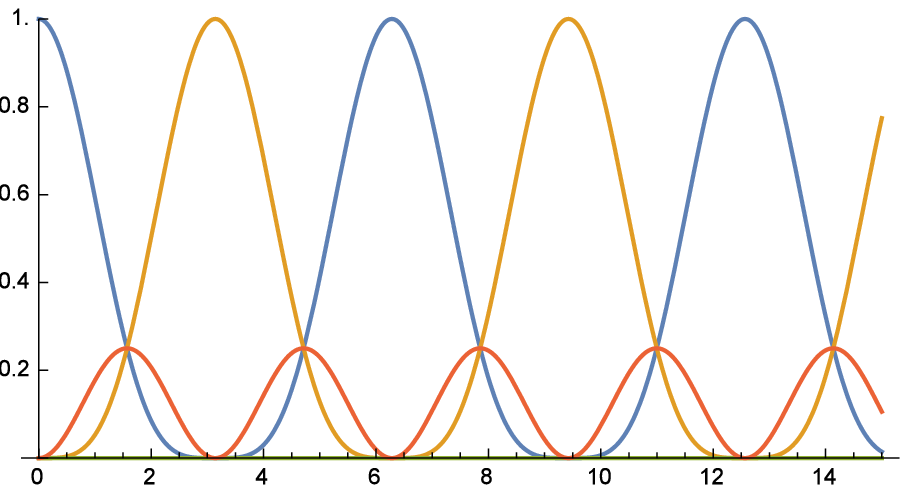}}
\put(94,.5){$\omega t$}
\put(8,44){$b_{\text{d1}}^{\rm neq}$}
\put(25,44){$b_{\rm d2}^{\rm neq}$}
\put(17,11){$b_{\rm qe}^{\rm neq}$}
\put(47,-7){\scalebox{1.5}{(d)}}
}
}
\end{picture}
\end{center}
\caption{Time dependence of matrix elements, $b_{\text{d1}}^{\rm neq}, b_{\text{d2}}^{\rm neq}, b_{\text{cf}}^{\rm neq}$ and $b_{\text{qe}}^{\rm neq}$, in dissipative and non-dissipative systems with scaled temperature, $(\beta J)^{-1}=0.7$. Parts (a), (b) and (c) show cases with a scaled dissipation rate of $\epsilon /\omega =1,~0.1$ and $0.01$, respectively. The dotdashed, dotted, dotdotdashed and dashed lines in parts (a), (b) and (c) represent the asymptotes of the $b_{\text{d1}}^{\rm neq}, b_{\text{cf}}^{\rm neq}, b_{\text{d2}}^{\rm neq}$ and $b_{\text{qe}}^{\rm neq}$ curves, respectively. In part (d)  which is the non-dissipation case ($\epsilon =0$), $b_{\text{cf}}^{\rm neq}$ vanishes identically.}\label{fig:01}
\end{figure}

\begin{figure}
\begin{center}
\unitlength 1mm
\begin{picture}(150,90)
\put(0,50){
\scalebox{0.7}{
\put(0,0){\includegraphics{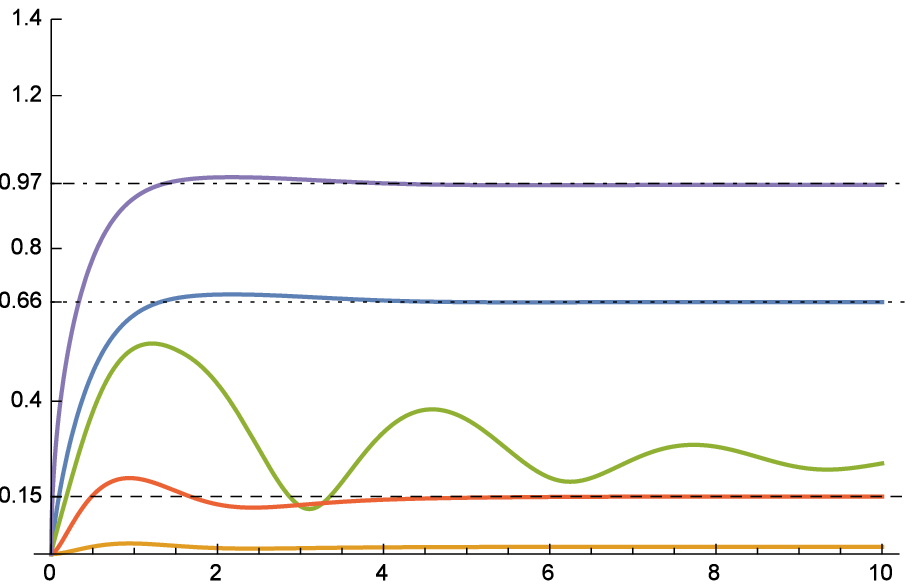}}
\put(94,.5){$\omega t$}
\put(17,25){$\hat{S}$}
\put(16,30){$S$}
\put(12,6){$b_{\text{qe}}^{\rm neq}$}
\put(15,12){$\hat{S}_{\text{qe}}$}
\put(17,43){$C$}
\put(47,-7){\scalebox{1.5}{(a)}}
}
}
\put(78,50){
\scalebox{0.7}{
\put(0,0){\includegraphics{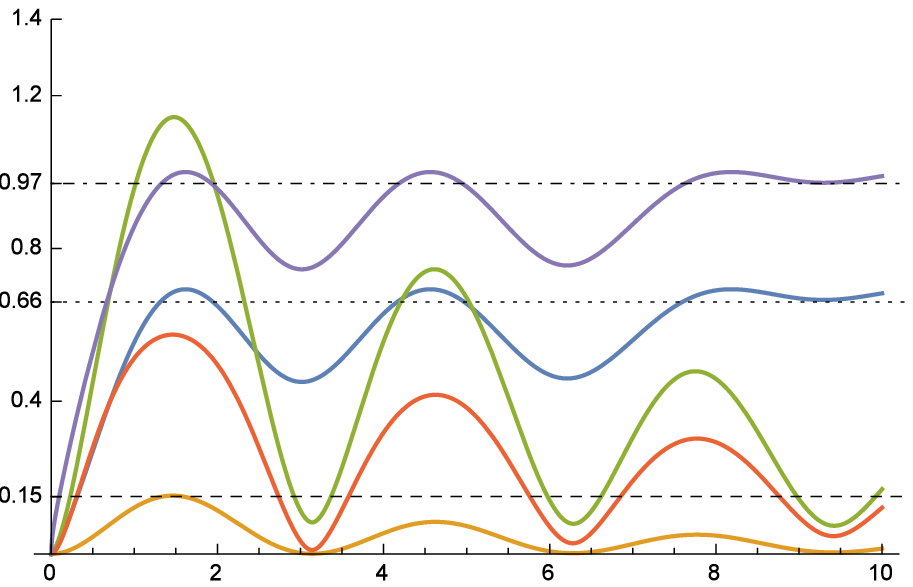}}
\put(94,.5){$\omega t$}
\put(16,49){$\hat{S}$}
\put(17,31.5){$S$}
\put(16,11){$b_{\text{qe}}^{\rm neq}$}
\put(16,21){$\hat{S}_{\text{qe}}$}
\put(16.5,38){$C$}
\put(47,-7){\scalebox{1.5}{(b)}}
}
}
\put(0,0){
\scalebox{0.7}{
\put(0,0){\includegraphics{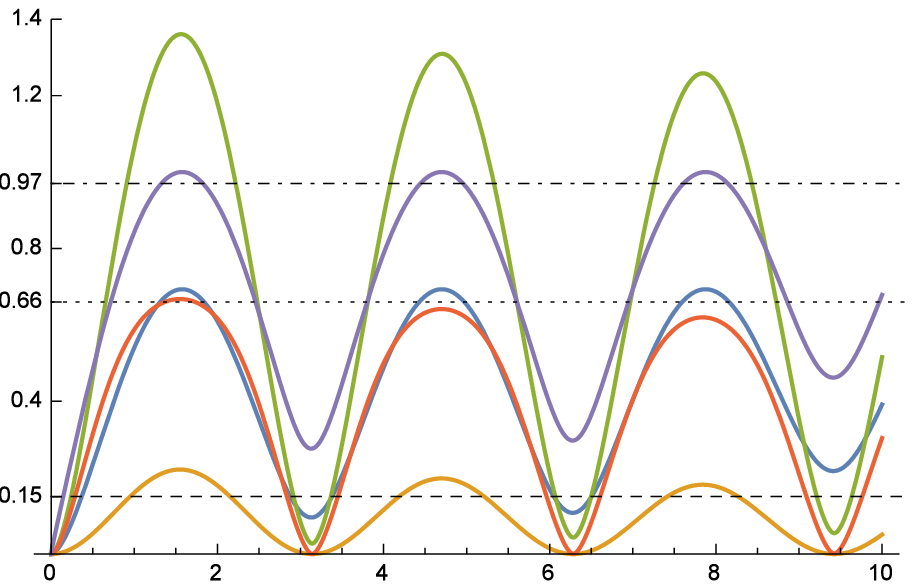}}
\put(94,.5){$\omega t$}
\put(17,52){$\hat{S}$}
\put(17,31.5){$S$}
\put(17,13.5){$b_{\text{qe}}^{\rm neq}$}
\put(16,25){$\hat{S}_{\text{qe}}$}
\put(16.5,43){$C$}
\put(47,-7){\scalebox{1.5}{(c)}}
}
}
\put(78,0){
\scalebox{0.7}{
\put(1.5,0){\includegraphics{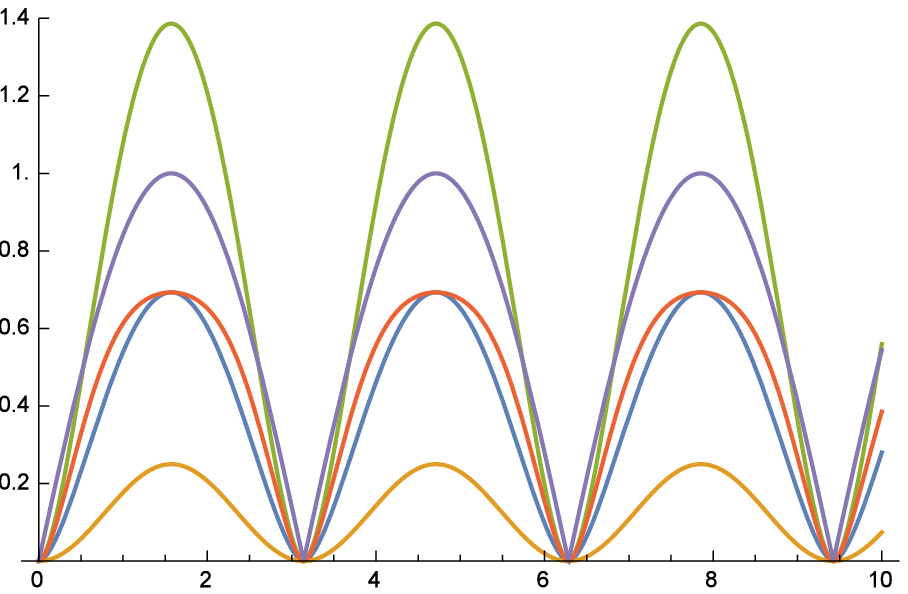}}
\put(95.5,.5){$\omega t$}
\put(17,53){$\hat{S}$}
\put(19,24){$S$}
\put(18,14.5){$b_{\text{qe}}^{\rm neq}$}
\put(18,32){$\hat{S}_{\text{qe}}$}
\put(17,44){$C$}
\put(47,-7){\scalebox{1.5}{(d)}}
}
}
\end{picture}
\end{center}
\caption{Time dependence of $C,~S,~\hat{S}$ and $\hat{S}_{\text{qe}}$, along with parameter $b_{\text{qe}}^{\rm neq}$, in dissipative and non-dissipative systems with $(\beta J)^{-1}=0.7$ and $\mu _BH/J=0.3$. Parts (a), (b) and (c) show cases with a scaled dissipation rate of $\epsilon /\omega =1,~0.1$ and $0.01$, respectively. The dotdashed, dotted, and dashed lines in parts (a), (b) and (c) represent the asymptotes of the $C,~S$ and $\hat{S}_{qe}$ curves, respectively. In (a), $\sim$, (c), $b_{\text{qe}}^{\rm neq}$ converge to the value $0.019\cdots$. In part (d) which is the non-dissipation case ($\epsilon =0$), all the curves have the same phase.}\label{fig:02}
\end{figure}

We are now ready to study the extended density matrix in the TFD double Hilbert space.
In terms of the expression similar to Eq. \eqref{eq:6}, we are then led to the renormalized extended density matrix, $\hat{\rho }_{\text{A}}^{\rm neq}$, as
\begin{align}
\hat{\rho}_{\text{A}}^{\rm neq} &=b_{\text{d1}}^{\rm neq}|+\rangle\langle+||\tilde{+}\rangle\langle\tilde{+}| +b_{\text{d2}}^{\rm neq}|-\rangle\langle-||\tilde{-}\rangle\langle\tilde{-}|\nonumber \\ &+b_{\text{cf}}^{\rm neq}\left(|+\rangle\langle-||\tilde{+}\rangle\langle\tilde{-}|+|-\rangle\langle+||\tilde{-}\rangle\langle\tilde{+}| \right)\nonumber \\ 
&+b_{\text{qe}}^{\rm neq}\left( |+\rangle\langle+||\tilde{-}\rangle\langle\tilde{-}|+|-\rangle\langle-||\tilde{+}\rangle\langle\tilde{+}|
\right), \label{eq:17}
\end{align}
where the matrix elements, $b_{\text{d1}}^{\rm neq}, b_{\text{d2}}^{\rm neq}, b_{\text{cf}}^{\rm neq}$ and $b_{\text{qe}}^{\rm neq}$, are respectively expressed as analytic functions of $H,~\epsilon,~t$ and $\beta$. However, their expressions are so complicated that we show their numerical results of several cases in Fig.~\ref{fig:01}.

The extended entanglement entropy, $\hat{S}:=-k_{\text{B}}\mathrm{Tr}_{\text{A}}\left[ \hat{\rho}_{\text{A}}^{\rm neq}\log \hat{\rho}_{\text{A}}^{\rm neq} \right] 
$, in the present case is then reduced to
\begin{equation}
\hat{S}=\hat{S}_{\rm cl}+\hat{S}_{\rm qe},
\label{eq:23}
\end{equation}
where
\begin{align}
\hat{S}_{\rm cl}&:=-k_{\text{B}}\left( \sqrt{4\left( b_{\text{cf}}^{\rm neq} \right) ^2+\left(b_{\text{d1}}^{\rm neq}-b_{\text{d2}}^{\rm neq}\right)^2} ~\mathrm{arccoth}\,
\frac{b_{\text{d1}}^{\rm neq}+b_{\text{d2}}^{\rm neq}}{\sqrt{4\left( b_{\text{cf}}^{\rm neq} \right) ^2+\left(b_{\text{d1}}^{\rm neq}-b_{\text{d2}}^{\rm neq}\right)^2}}\right.\notag \\
&\hspace{3mm}\left. +\frac{b_{\text{d1}}^{\rm neq}+b_{\text{d2}}^{\rm neq}}{2}\log \left(b_{\text{d1}}^{\rm neq}b_{\text{d2}}^{\rm neq}-\left( b_{\text{cf}}^{\rm neq} \right) ^2\right)\right),
\label{eq:23a}
\end{align}
and
\begin{align}
\hat{S}_{\rm qe}:=-2k_{\text{B}}b_{\text{qe}}^{\rm neq}\log b_{\text{qe}}^{\rm neq},
\label{eq:23b}
\end{align}
respectively. In Eqs.~\eqref{eq:23}, \eqref{eq:23a} and \eqref{eq:23b}, the expressions of $\hat{S}$, the classical and thermal fluctuation parts, $\hat{S}_{\rm cl}$, and the quantum entanglement part, $\hat{S}_{\rm qe}$, also comprise analytic functions of $ t, \beta, \epsilon$ and $\omega$, respectively, however, their expressions are quite complicated.
So, we show the numerical behaviour of $C,~S,~\hat{S},~\hat{S}_{\text{qe}}$ and $b_{\text{qe}}^{\rm neq}$ for a few cases in Figs.~\ref{fig:02}(a)$-$(c) (in units of $k_{\text{B}}=1$).
These figures exhibit that, at $t \to \infty $, $\hat{S},~\hat{S}_{\text{qe}}$ and $b_{\text{qe}}^{\rm neq}$ converge to the values, $0.277\cdots,~0.150\cdots$ and $0.019\cdots$, respectively. The traditional entanglement entropy, $S$, consequently becomes larger than the extended entanglement entropies, $\hat{S}$ and $\hat{S}_{\rm qe}$, at $t \to \infty $.
 These results suggest that these numerical behaviour of $C,~S,~\hat{S},~\hat{S}_{\text{qe}}$ and $b_{\text{qe}}^{\rm neq}$ is resemblant to the one in the case of no external field \cite{Nakagawa}.


For non-dissipative systems, $\hat{S}$ in Eq.~\eqref{eq:23} and $\hat{S}_{\rm qe}$ in Eq.~\eqref{eq:23b} reduce to
\begin{align}
\hat{S}=-k_{\text{B}}&\left( \cos ^4 \frac{\omega t}{2}\cdot \log \left( \cos ^4 \frac{\omega t}{2} \right) +  \sin ^4 \frac{\omega t}{2}\cdot \log \left( \sin ^4 \frac{\omega t}{2} \right) \right.\nonumber \\ 
&\left.+ \frac{1}{2} \sin ^2 \omega t\cdot \log \left( \frac{\sin ^2 \omega t}{4} \right)\right),
\label{eq:24}
\end{align}
and 
\begin{align}
\hat{S}_{\rm qe}=- \frac{k_{\text{B}}}{2} \sin ^2 \omega t\cdot \log \left( \frac{\sin ^2 \omega t}{4} \right),
\label{eq:25}
\end{align}
respectively, at $\epsilon =0$. As can transparently be visualized from Eqs. \eqref{eq:24} and \eqref{eq:25}, the dependence of $\hat{S}$ and $\hat{S}_{\rm qe}$ on both $\beta $ and $H$ disappear at $\epsilon =0$ and this result assorts with the case of no external field. The time dependence of $\hat{S}$ and $\hat{S}_{\rm qe}$ at $\epsilon =0$ is exhibited in Fig.~\ref{fig:01}(d) (in units of $k_{\text{B}}=1$). It is apparent in this figure that all the curves ($C,~S,~\hat{S},~\hat{S}_{\text{qe}}$ and $b_{\text{qe}}$) showing the entanglement have the same phase, however, their amplitudes differ.
This result is also similar to the case of no external field \cite{Nakagawa}. 

\section{Conclusions }
\label{sec:4}

In this communication, we have examined the extended density matrices and the extended entanglement entropies of finite-spin systems with and without non-competitive external fields in both the equilibrium and non-equilibrium cases, based upon the TFD formulation. 
For the equilibrium system, the extended density matrix elements are derived using TFD, and the low and high temperature limits are calculated explicitly. The results show that the effect of the non-competitive external field is significant for the temperature dependence of the extended density matrix. For the non-equilibrium systems, based upon the extended density matrix, $C,~S,~\hat{S},~\hat{S}_{\text{qe}}$ and $b_{\text{qe}}^{\rm neq}$, are computed using TFD, and summarized in Figs. \ref{fig:01} and \ref{fig:02}. 
From these results, we can deduce the following:
\begin{description}
 \item{1)} The each shapes of the curves of $b_{\text{d1}}^{\rm neq},~b_{\text{d2}}^{\rm neq}~\mbox{and}~b_{\text{qe}}^{\rm neq}$ in Fig. \ref{fig:01}(b) are similar to those of Fig. 4 in Ref. \cite{Hashizume}, except that the curve of $b_{\text{cf}}^{\rm neq}$ is compartively wavy.
 \item{2)} The each behaviours of $b_{\text{d1}}^{\rm neq},~b_{\text{d2}}^{\rm neq}~,b_{\text{cf}}^{\rm neq}~\mbox{and}~b_{\text{qe}}^{\rm neq}$ in Figs. \ref{fig:01}(a), (c) and (d) are unparalleled. 
 \item{3)} The each shapes of the curves of $C,~S,~\hat{S},~\hat{S}_{\text{qe}}$ and $b_{\text{qe}}^{\rm neq}$ in Figs. \ref{fig:02}(a), (b), (c) and (d) are also similar to those of Figs. 1 (a), (b), (c) and (d) in Ref. \cite{Nakagawa}.
\end{description}
As a consequence, the non-competitive external field has been found to have little effect on $C,~S,~\hat{S},~\hat{S}_{\text{qe}}$ and $b_{\text{qe}}^{\rm neq}$, and modest effect on $b_{\text{cf}}^{\rm neq}$. So another thing to keep in mind is to examine another spin system with competitive external field using the TFD formulation. However, the dissipative dynamics of entanglement in frustrated spin systems is more complicated, and is under consideration.

We have anyhow succeeded in elucidating the extended density matrices and the entanglement entropies of finite-spin systems with and without the non-competitive external fields which are constructed in both the equilibrium and non-equilibrium cases, based upon the general TFD algorithm.
Let us conclude by emphasizing that the new TFD-based method enables us to clearly distinguish between the various states of quantum systems.



\end{document}